\documentclass{article}

\usepackage[accepted]{icml2020}

\usepackage[utf8]{inputenc} 
\usepackage[T1]{fontenc}    
\usepackage{hyperref}       
\usepackage{url}            
\usepackage{booktabs}       
\usepackage{amsfonts}       
\usepackage{nicefrac}       
\usepackage{microtype}      
\usepackage{multirow}
\usepackage{graphicx}
\usepackage{algorithm}
\usepackage{amsmath}
\usepackage{amsfonts}
\usepackage{amssymb}
\usepackage{subfig}

\newcommand{\beginsupplement}{
    \setcounter{table}{0}
    \renewcommand{\thetable}{S\arabic{table}}
    \setcounter{figure}{0}
    \renewcommand{\thefigure}{S\arabic{figure}}}

\icmltitlerunning{Learning Quadratic Games on  Networks}

\begin{document}

\twocolumn[
\icmltitle{Learning Quadratic Games {on} Networks}



\icmlsetsymbol{equal}{*}

\begin{icmlauthorlist}
\icmlauthor{Yan Leng}{1}
\icmlauthor{Xiaowen Dong}{2}
\icmlauthor{Junfeng Wu}{3}
\icmlauthor{Alex Pentland}{4}
\end{icmlauthorlist}

\icmlaffiliation{1}{McCombs School of Business, The University of Texas at Austin, Austin, TX, USA}
\icmlaffiliation{2}{Department of Engineering Science, University of Oxford, Oxford, UK}
\icmlaffiliation{3}{College of Control Science and Engineering, Zhejiang University, Hangzhou, China}
\icmlaffiliation{4}{Media Lab, Massachusetts Institute of Technology, Cambridge, MA, USA}

\icmlcorrespondingauthor{Yan Leng}{yan.leng@mccombs.utexas.edu}

\icmlkeywords{Machine Learning, ICML}

\vskip 0.3in
]



\printAffiliationsAndNotice{}  
    \begin{abstract}
        Individuals, or organizations, cooperate with or compete against one another in a wide range of practical situations.
        Such strategic interactions are often modeled as games played on networks, where an individual's payoff depends not only on her action but also on that of her neighbors. The current literature has largely focused on analyzing the characteristics of network games in the scenario where the structure of the network, which is represented by a graph, is known beforehand. It is often the case, however, that the actions of the players are readily observable while the underlying interaction network remains hidden.
        In this paper, we propose two novel frameworks for learning, from the observations on individual actions, 
        {network games with linear-quadratic payoffs, and in particular the structure of the interaction network.} Our frameworks are based on the Nash equilibrium of such games and involve solving a joint optimization problem for the graph structure and the individual marginal benefits. 
        Both synthetic and real-world experiments demonstrate the effectiveness of the proposed frameworks, which
        have theoretical as well as practical implications for understanding strategic interactions in a network environment.
    \end{abstract}

    \section{Introduction}
    We live in an increasingly connected society. First studied by the American sociologist Stanley Milgram via his 1960s experiments and later popularized by John Guare's 1990 eponymous play, the theory of "six degrees of separation" has been recently re-analyzed on the social networking site Facebook, only to find out that any pair of Facebook users can actually be connected via approximately three and a half other ones \cite{backstrom2012four}. Individuals, unsurprisingly, are not merely connected; their decisions and actions often influence the ones around them. Indeed, Christakis and Fowler \cite{christakis2009connected} have found in a series of studies that, one's emotion, health habit, and political opinion can affect individuals who are as far as three degrees of separation in her social circle. Furthermore, such influence on the decision-making process may take place via either explicit \cite{aral2009distinguishing,Leng2018Rippling} or implicit interactions \cite{bandura1977social,dong2018social}.
    
    To study the decision-making of a group of interacting agents, recent literature in economics has increasingly focused on the modeling of such interactions as games played on networks \cite{jackson2014games,bramoull2016games}. The underlying assumption in this setting is that, in a game played by a group of players who form a social network, the payoff of a player depends on her action, e.g., an effort made to achieve a specific task, as well as that of her neighbors in the network. Two types of actions have been studied in the literature, i.e., strategic complements and strategic substitutes. In the former case, 
    one's action increases her neighbors' incentives for action, e.g., students putting an effort together into a joint assignment or firms working on a collaborative research project \cite{goyal2001r}. In the latter case, however, the situation is the opposite,
    such as the scenarios of firms competing on market prices or individuals on local public goods \cite{bramoulle2007public}.
    
    In a network game, the underlying structure of the network carries critical information and dictates the behavior and actions of the players. Typically, graphs are used as mathematical tools to represent the structure of these networks, and the current literature in this area has predominately focused on studying the characteristics of games on known or predefined graphs \cite{ballester2006who,bramoulle2014strategic,galeotti2017targeting}. However, it is increasingly common that while ample observations on the actions of the agents are available, the underlying complex relationships among them, which may be captured by an interaction network, remain mostly hidden due to cost in observation or privacy concern.
    In this case, the network needs to be estimated to better understand the present and predict the future actions of these agents. The primary goal of this paper is, therefore, to study the problem of learning, given the observations on the actions of the agents, a graph structure that best explains the observed actions in the setting of a network game. 
    
    Such a problem, generally speaking, may be thought of as an instance of the ones of learning relationships, often in the form of graph structures, from observations made on a set of data entities.
    Classical approaches from the machine learning and signal processing communities tackle this problem by building statistical models (e.g., probabilistic graphical models \cite{Koller09, Friedman08}), physically-motivated models (e.g., diffusion processes on networks \cite{GomezRodriguez_2010, Gomez-Rodriguez20011}), or more recently signal processing models \cite{Dong19, Mateos19}. These approaches, however, do not take into account the game-theoretic aspect of the decision-making of players in a network environment. 
    
    In the computer science literature, network games are known as graphical games \cite{kearns2001graphical}, and there have been a few studies recently on learning the games from observed action data.
    For example, the works in \citet{irfan2014influence,honorio2015learning,ghoshal16,ghoshal2017learning} have proposed to learn graphical games by observing actions from linear influence games with linear influence functions, where \citet{ghoshal2018learning} has considered polymatrix games with pairwise matrix payoff functions. 
    The work in \citet{garg2016learning} has proposed to learn potential games on tree-structured networks of symmetric interactions. These conditions have been relaxed in \citet{garg2017local} where the authors have
    studied aggregative games where a player's payoff is convex and Lipschitz in an aggregate of their neighbors' actions defined via a local aggregator function. 
    All these works, however, either consider a binary or a finite discrete action space, which may be restrictive in certain practical scenarios where actions take continuous values.
    {Very recently, \citet{Barik19} has considered learning continuous-action graphical games, which is similar in spirit to our study albeit under a slightly different action (which is budgeted) and payoff setting.}
    
    
    In this study, we focus on learning games with linear-quadratic payoffs \cite{ballester2006who,bramoulle2014strategic,acemoglu2015networks,galeotti2017targeting}.
    We propose a learning framework 
    where, {given the Nash equilibrium action of the games,} we jointly infer the graph structure that represents the interaction network as well as the individual marginal benefits. We further develop a second framework by considering the homophilous effect of individual marginal benefits in the interaction network. The first framework involves solving a convex optimization problem, while the second leads to a non-convex one for which we develop an algorithm based on alternating minimization. We test the performance of the proposed algorithms in inferring graph structures for network games and show that it is superior to the baseline approaches of sample correlation and regularized graphical Lasso \cite{lake2010discovering}, albeit developed for slightly different learning settings.
    
    The main contributions of the paper are as follows. First, the proposed learning frameworks, to the best of our knowledge, are the first to 
    {address the problem of learning the graph structure of the broad class of network games with linear-quadratic payoffs and continuous actions.}
    {Second, our framework also allows for the inference of marginal benefits of the players which permits a range of applications such as target interventions.}
    Third, we analyze several factors in the quadratic games that affect the learning performance, such as the strength of strategic complements or substitutes, the topological characteristics of the networks, and the homophilous effect of individual marginal benefits.
    Overall, our paper constitutes a theoretical contribution to the studies of network games and may shed light on the understanding of strategic interactions in a wide range of practical scenarios, including business, education, governance, and technology adoption.

    \section{Network Games of Linear-Quadratic Payoffs}
    \label{game}
    {Consider a { weighted} network of $N$ individuals represented by a graph $\mathcal{G}(\mathcal{V}, \mathcal{E})$, where $\mathcal{V}$ and $\mathcal{E}$ denote the node and edge sets, respectively. For any pair of individuals $i$ and $j$, {$G_{ij} = G_{ji} > 0$} if $(i,j) \in \mathcal{E}$ and $G_{ij} = G_{ji} = 0$ otherwise, where $G_{ij}$ is the $ij$-th entry of the adjacency matrix $\mathbf{G}$. }
    In a network game of linear-quadratic payoffs, an individual $i$ chooses her action $a_i$ to maximize her payoff, $u_i$, which has the following form \cite{ballester2006who,bramoulle2014strategic,galeotti2017targeting,acemoglu2015networks}:
    \begin{equation}
    u_i = b_i a_i -\frac{1}{2}a_i^2 + \beta a_i \sum_{j \in \mathcal{V}} G_{ij} a_j.
    \label{utility}
    \end{equation}
    In Eq.~(\ref{utility}), the first term is contributed by $i$'s own action where the parameter $b_i$ is called the marginal benefit, and the third term comes from the peer effect weighted by the actions of her neighbors. The parameter $\beta$ captures the nature and the strength of such peer effect: if $\beta > 0$, actions are called strategic complements; and if $\beta < 0$, actions are called strategic substitutes.
    
    {The quadratic game with payoff function in Eq.~(\ref{utility}) represents a broad class of games that have been extensively studied in the literature, and has a number of desirable properties. First, it naturally allows for continuous actions (i.e., $a_i$ is considered to be continuous); second, it can be used for modelling games of both strategic complements and substitutes, {i.e., positive and negative spillover effect;} 
    third, it may also be used to approximate games with complex non-linear payoffs. For these reasons, games of linear-quadratic payoffs have been used to analyse crime activity, educational outcome, firm cooperation, and urban dynamics just to name a few \cite{jackson2014games}.}
    
    {One important advantage of the game in Eq.~(\ref{utility}) is that it allows for an explicit
    solution for equilibrium behavior as a function of the network. To see this,} 
    let us define the vectorial forms $\mathbf{a} = [a_1,a_2,\cdots,a_N]^T$, $\mathbf{b} = [b_1,b_2,\cdots,b_N]^T$, and $\mathbf{u} = [u_1,u_2,\cdots,u_N]^T$, where we use the convention that the subscript $i$ indicates the $i$-th entry of the vector. Taking the first-order derivative of the payoff $u_i$ with respect to the action $a_i$ in Eq.~(\ref{utility}), we have:
    \begin{equation}
    \frac{\partial u_i}{\partial a_i} = b_i - a_i + \beta (\mathbf{G} \mathbf{a})_i. 
    \label{b}
    \end{equation}

    Combining Eq.~(\ref{b}) for all $i$, it is clear that the following relationship holds, as pointed out in \citet{ballester2006who}, for any (pure strategy) Nash equilibrium action $\mathbf{a}$: 
    \begin{equation}
    \left( \mathbf{I} - \beta \mathbf{G} \right)\mathbf{a} = \mathbf{b},
    \label{b_equi}
    \end{equation}
    hence
    \begin{equation}
    \mathbf{a} = \left( \mathbf{I} - \beta \mathbf{G} \right)^{-1}\mathbf{b},
    \label{equi}
    \end{equation}
    where $\mathbf{I} \in \mathbb{R}^{N \times N}$ is the identity matrix.
    We adopt the critical assumption that the spectral radius of the matrix $\beta \mathbf{G}$, denoted by $\rho(\beta \mathbf{G})$, is less than 1, which guarantees the inversion of Eq.~(\ref{equi}). Furthermore, as proved in \citet{ballester2006who}, this assumption also ensures the uniqueness and stability of the Nash equilibrium action $\mathbf{a}$. 
    

    {The equilibrium action $\mathbf{a}$ can be rewritten as $\mathbf{a} = \sum_{p=0}^{\infty} \beta^p \mathbf{G}^p \mathbf{b}$, and therefore has the following interpretations. If $\mathbf{b}$ is the all-one vector, then each entry of $\mathbf{a}$ is the Katz-Bonacich centrality \cite{Katz53, Bonacich87} of the corresponding node, i.e., the number of walks of any length $p$ originated from that node discounted exponentially by $\beta$. As pointed out in \citet{jackson2014games}, interestingly, this means despite the local neighborhood relationship in Eq.~(\ref{utility}) the payoff interdependency actually spreads indirectly throughout the network.}
    On the other hand, the formulation of Eq.~(\ref{equi}) can also be interpreted as computing steady-state opinions in studying opinion dynamics under a linear DeGroot model \cite{DeGroot74} and has been used in works on social network sensing \cite{Wai16}.
    
    {From a different perspective, notice that $\mathbf{G}$ is a real and symmetric matrix hence has the following eigendecomposition: $\mathbf{G} = \boldsymbol{\chi} \boldsymbol{\Lambda} \boldsymbol{\chi}^T$. Plugging this into Eq.~(\ref{equi}), the equilibrium action $\mathbf{a}$ can then be rewritten as $\mathbf{a} = \boldsymbol{\chi} (\mathbf{I} - \beta \boldsymbol{\Lambda})^{-1} \boldsymbol{\chi}^T \mathbf{b}$. Treating the marginal benefit $\mathbf{b}$ as a signal defined on the node set of the graph, the operation $\boldsymbol{\chi}^T \mathbf{b}$ can be interpreted as a Fourier-like transform for $\mathbf{b}$ according to the graph signal processing literature \cite{Ortega18}. Given that the eigenvector associated with the largest/smallest eigenvalue of $\mathbf{G}$ is the most smooth/non-smooth hence corresponds to low-/high-frequency signal on the graph, the action $\mathbf{a}$ can thus be interpreted as a low-pass filtered version of $\mathbf{b}$ for $\beta>0$, and a high-pass filtered version of it for $\beta<0$. This matches our intuition that equilibrium action tends to be smooth on the interaction network for the case of strategic complements, and non-smooth for strategic substitutes.} 

    \section{Learning Games with Independent Marginal Benefits}
    \label{algo1}
    Given the graph with an adjacency matrix $\mathbf{G}$, the marginal benefits $\mathbf{b}$, and the parameter $\beta$, Eq.~(\ref{equi}) provides a way of computing $\mathbf{a}$, the Nash equilibrium action of the players. 
    The graph structure, in many cases, can be naturally chosen from the application domain, such as a social or business network. 
    However, these natural choices of graphs may not necessarily describe well the strategic interactions between the players, and a natural graph might not be easy to define at all in some applications. Compared to the underlying relationships captured by $\mathbf{G}$, it is often easier to observe the individual actions $\mathbf{a}$, such as the amount of effort committed by students in a joint course project, or the strategic moves made by firms in an industrial setting.
    In these cases, given the actions and the dependencies described in Eq.~(\ref{utility}), it is therefore of considerable interest to infer the structure of the graph on which the game is played, hence revealing the hidden relationships between the players.

    \subsection{Learning Framework}
    \label{sec:form1}
    We consider $N$ players, connected by a fixed interaction network $\mathbf{G}$, playing $K$ different and independent games in each of which their payoffs depend not only on their own actions but also that of their neighbors\footnote{{The setting of $K$ games of scalar actions could also be interpreted as one game with a $K$-dimensional action, in which case the assumption is that each dimension of the player's action satisfies the equilibrium condition.}}. Let us define the marginal benefits for these $K$ games as $\mathbf{B} = [\mathbf{b}^{(1)}, \mathbf{b}^{(2)}, \cdots, \mathbf{b}^{(k)}] \in \mathbb{R}^{N \times K}$, where each column of $\mathbf{B}$ is the marginal benefit vector for one game, and the corresponding actions of the players as $\mathbf{A} = [\mathbf{a}^{(1)}, \mathbf{a}^{(2)}, \cdots, \mathbf{a}^{(K)}] \in \mathbb{R}^{N \times K}$. 
    We first consider in this section the case where, for each game, the marginal benefits of individual players follow independent and identical Gaussian distributions, and then address in Section~\ref{algo2} the dependent case.
    The parameter that captures the strength of the network effect, $\beta$, can be either positive or negative, corresponding to strategic complements and strategic substitutes, respectively. {Notice that we assume a fixed beta for all games in the current setting, which implies that: 1) the relationship between a player and her neighbors remains the same for all players (either complements or substitutes); 2) the contribution to the individual payoff from the neighbors is scaled by a beta for all players. A more flexible setting for the $\beta$ parameter could be an interesting future direction.}
    
    
    {Given the observed actions $\mathbf{A}$ and the parameter $\beta$, the goal is to infer a graph structure $\mathbf{G}$ {as well as the marginal benefits $\mathbf{B}$, which best explain $\mathbf{A}$ in terms of the relationship in Eq.~(\ref{b_equi}).}}
    To this end, we propose the following joint optimization problem of $\mathbf{G}$ and $\mathbf{B}$: 
    \begin{equation}
    \begin{split}
    \underset{\mathbf{G}, \mathbf{B}}{\text{min}}~~& f(\mathbf{G}, \mathbf{B}) \\
    = & ||(\mathbf{I} - \beta  \mathbf{G} )\mathbf{A} - \mathbf{B}||_F^2 + \theta_1||\mathbf{G}||_F^2 + \theta_2||\mathbf{B}||_F^2, \\
    \text{s.t.}~~& G_{ij} = G_{ji},~G_{ij} \geq 0,~G_{ii} = 0~~\text{for}~~\forall i,j \in \mathcal{V}, \\
    & ||\mathbf{G}||_1 = N,
    \end{split}
    \label{obj_min1}
    \end{equation}
    where { $||\cdot||_F$ and $||\cdot||_1$ denote the  Frobenius norm and element-wise $L^1$-norm of a matrix,} respectively, and $\theta_1$ and $\theta_2$ are two non-negative regularization parameters. 
    The first line of constraints ensures that $\mathbf{G}$ is a valid adjacency matrix, and the second constraint (the constraint on the $L^1$-norm) fixes the volume of the graph and permits to avoid trivial solutions. {Without loss of generality the volume is chosen to be $N$.}
    It is clear that, in the problem of Eq.~(\ref{obj_min1}), we aim at a joint inference of the graph structure $\mathbf{G}$ and the marginal benefits $\mathbf{B}$, such that the observed actions $\mathbf{A}$ are close to the Nash Equilibria of the $K$ games played on the graph.
    The Frobenius norm on $\mathbf{G}$ is added as a penalty term to control the distribution of the edge weights of the learned graph (the off-diagonal entries of $\mathbf{G}$)\footnote{Similar constraints have been adopted in \citet{Hu15,Dong16} for graph inference.}, which, together with the $L^1$-norm constraint, bears similarity to the linear combination of $L^1$ and $L^2$ penalties in an elastic net regularization \cite{Zou05}.
    
    {The effectiveness of the formulation in Eq.~(\ref{obj_min1}) depends on $\rho(\beta G)$. To see this, notice that under the assumption that $\mathbf{b}$ is IID Gaussian, the equilibrium action $\mathbf{a}$ follows a Gaussian distribution with covariance $(\mathbf{I} - \beta  \mathbf{G})^{-2}$. If $\rho(\beta G)$ is close to zero, then $\mathbf{a}$ is almost independent from $G$, and it would be difficult to infer $\mathbf{G}$ from $\mathbf{a}$ in this scenario. On the other hand, if $\rho(\beta G)$ is close to one, the covariance is dominated by the eigenvector associated with the largest (when $\beta>0$) or smallest (when $\beta<0$) eigenvalue of $\mathbf{G}$. In this case, the action $\mathbf{a}$ clearly contains information about $\mathbf{G}$ which facilitates learning.}
    
    {It can be seen from Eq.~(\ref{obj_min1}) that, comparing to the work in \citet{Barik19} which learns the neighbors for each player separately via a regression framework, our approach learns the network at once by solving a single optimization problem. This difference is analogous to that between the neighborhood selection of \citet{meinshausen2006high} and the graphical Lasso of \citet{Friedman08} in the context of covariance estimation.}
    
    \subsection{Learning Algorithm}
    \label{graph_non_smooth_alg}
    Given the non-negativity of $G_{ij}$, we can re-write the constraint: $||\mathbf{G}||_1 = \mathbf{1}^T \mathbf{G} \mathbf{1} = N$, where $\mathbf{1} \in \mathbb{R}^N$ is the all-one vector. The constraints in Eq.~(\ref{obj_min1}) therefore form a convex set.
    The problem of Eq.~(\ref{obj_min1}) is thus a quadratic program jointly convex in $\mathbf{B}$ and $\mathbf{G}$,
    and can be solved efficiently via the interior point methods \cite{boyd2004convex}. In our experiments, we solve the problem of Eq.~(\ref{obj_min1}) using the Python software package CVXOPT \cite{andersen2018cvxopt}. In case of graphs of very large number of vertices, we can instead use operator splitting methods, e.g., alternating direction method of multipliers (ADMM) \cite{boyd2011distributed}, to find a solution. The algorithm is summarized in Algorithm~\ref{alg:non-smooth}.
    

    \begin{algorithm}[h]
        \caption{Learning Games with Independent Marginal Benefits}
        \label{alg:non-smooth}
        \begin{algorithmic}
            \STATE {\bfseries Input:} Actions $\mathbf{A}\in \mathbb{R}^{N\times K}$ for $K$ games, $\mathbf{\beta}$, $\theta_1$, $\theta_2$
            \STATE {\bfseries  Output:} Network $\mathbf{G} \in \mathbb{R}^{N\times N}$, marginal benefits $\mathbf{B} \in \mathbb{R}^{N\times K}$ for $K$ games
            \STATE Solve for $\mathbf{G}$ and $\mathbf{B}$ in Eq.~(\ref{obj_min1})
            \STATE {\bfseries return: } $\mathbf{G}$, $\mathbf{B}$  
        \end{algorithmic}
    \end{algorithm}

    \section{Learning Games with Homophilous Marginal Benefits}
    \label{algo2}
    A large number of studies in the literature of social sciences and economics have analyzed the phenomenon of homophily in social networks, which describes that individuals tend to associate and form ties with those that are similar to themselves \cite{mcpherson2001birds,jackson2010social}. Since the marginal benefit vector $\mathbf{b}$ in each game can be thought of as the individual preferences toward a particular action, they may contribute, in the presence of the homophily effect, to the formation of the interaction network on which the game is played.
    The second formulation in our paper is, therefore, to address the problem of learning {games} with such homophilous marginal benefits.

    \subsection{Learning Framework}
    \label{sec:form2}
    The homophily effect that is present in the marginal benefit vector $\mathbf{b}$ implies that $\mathbf{b}$ as a signal defined on the graph is relatively smooth, in the sense that nodes that are connected share similar marginal benefits. This may be quantified by the so-called Laplacian quadratic form on the graph:
    \begin{equation}
    \mathbf{b}^T \mathbf{L} \mathbf{b} = \frac{1}{2} \sum_{i,j \in \mathcal{V}} G_{ij} \left(b_i-b_j\right)^2,
    \label{eq:lapquad}
    \end{equation}
    where $\mathbf{L} = \text{diag}(\sum_{j \in \mathcal{V}} G_{ij}) - \mathbf{G}$ is the unnormalized (combinatorial) graph Laplacian matrix \cite{Chung97}. We therefore propose to replace the norm on $\mathbf{B}$ with this measure in the objective function of Eq.~(\ref{obj_min1}) to promote homophilous marginal benefits. This essentially assumes that the marginal benefits follow a multivariate Gaussian distribution with the precision matrix being the graph Laplacian. This leads to the following optimization problem:
    \begin{equation}
    \begin{split}
    \underset{\mathbf{G},\mathbf{B}}{\text{min}}~~& h(\mathbf{G}, \mathbf{B}) \\
    = & ||(\mathbf{I} - \beta  \mathbf{G} )\mathbf{A} - \mathbf{B}||_F^2 + \theta_1||\mathbf{G}||_F^2 +  \theta_2~\text{tr}(\mathbf{B}^T \mathbf{L} \mathbf{B}), \\
    \text{s.t.}~~& G_{ij} = G_{ji},~G_{ij} \geq 0,~G_{ii} = 0~~\text{for}~~\forall i,j \in \mathcal{V}, \\
    & ||\mathbf{G}||_1 = N, \\
    & \mathbf{L} = \text{diag}(\sum_{j \in \mathcal{V}} G_{ij}) -  \mathbf{G},
    \end{split}
    \label{homo_obj}
    \end{equation}
     { where $\text{tr}(\cdot)$  denotes the trace operator.} The third term in the objective is the sum of the Laplacian quadratic form for all the columns in $\mathbf{B}$, and the third constraint comes from the definition of the graph Laplacian $\mathbf{L}$. Like in Eq.~(\ref{obj_min1}), $\theta_1$ and $\theta_2$ are two non-negative regularization parameters.
    The problem of Eq.~(\ref{homo_obj}) is similar to that of Eq.~(\ref{obj_min1}), but with a different assumption that there exists the effect of homophily in the marginal benefits $\mathbf{b}$, whose strength is controlled by the regularization parameter $\theta_2$, i.e., a larger $\theta_2$ favors a stronger homophily effect, and vice versa.

    \subsection{Learning Algorithm}
    \label{graph_smooth_alg}
    Unlike the problem of Eq.~(\ref{obj_min1}), the problem of Eq.~(\ref{homo_obj}) is not jointly convex in $\mathbf{G}$ and $\mathbf{B}$ due to the third term in the objective function. We, therefore, adopt an alternating minimization scheme to optimize for the graph structure $\mathbf{G}$ and the marginal benefits $\mathbf{B}$ where, at each step, we solve for one variable while fixing the other. 
    
    Given $\mathbf{B}$, we first solve for $\mathbf{G}$ in Eq.~(\ref{homo_obj}). 
    The constraints on $\mathbf{G}$ in Eq.~(\ref{homo_obj}) are the same as that in Eq.~(\ref{obj_min1}) and thus convex. Since $\theta_1 \geq 0$ and $\theta_2 \geq 0$, fixing $\mathbf{B}$ and solving for $\mathbf{G}$ results in a strongly convex objective, and consequently the problem admits a unique solution.
    We again solve this convex quadratic program using the package CVXOPT.
    Next, we fix $\mathbf{G}$ and solve for $\mathbf{B}$ in Eq.~(\ref{homo_obj}). 
    By fixing $\mathbf{G}$, Eq.~(\ref{homo_obj}) becomes an unconstrained convex quadratic program, and thus admits a closed-form solution which can be obtained by setting the derivative to zero:
    \begin{equation}
    \frac{ \partial h(\mathbf{G}, \mathbf{B})}{\partial \mathbf{B}}= -2 \big((\mathbf{I} - \beta \mathbf{G}) \mathbf{A} - \mathbf{B} \big) + 2\theta_2 \mathbf{L} \mathbf{B} = \mathbf{0}, 
    \end{equation}
    hence
    \begin{equation}
    \mathbf{B} = (\mathbf{I} + \theta_2 \mathbf{L} )^{-1} (\mathbf{I} - \beta \mathbf{G})\mathbf{A}.
    \end{equation}
    
    
    We iterative between the two steps until either the change in the objective $h(\mathbf{G}, \mathbf{B})$ is smaller than $10^{-4}$, or a maximum number of iterations has been reached. 
    {This strategy is called block coordinate descent (BCD) and, since both subproblems are strongly convex, is guaranteed to converge to a local minimum (see Proposition 2.7.1 in \citet{Bertsekas95}).}
    The complete algorithm is summarized in Algorithm~\ref{alg:smooth}.
    
    \begin{algorithm}[h]
        \caption{Learning Games with Homophilous Marginal Benefits}\label{alg:smooth}
        \begin{algorithmic}
            \STATE \textbf{Input:} Actions $\mathbf{A}\in \mathbb{R}^{N\times K}$ for $K$ games, $\mathbf{\beta}$, $\theta_1$, $\theta_2$
            \STATE \textbf{Output:} Network $\mathbf{G} \in \mathbb{R}^{N\times N}$, marginal benefits $\mathbf{B} \in \mathbb{R}^{N\times K}$ for $K$ games
            \STATE \textbf{Initialize:} $\mathbf{B}_0(:,k) \sim \mathcal{N}(\mathbf{0}, { \mathbf{I}}$) \text{for} $k=1,\cdots, K$, $t=1$, $\Delta=1$
            \IF{$\Delta \geq 10^{-4}$ and $t \leq $ \# iterations}
            \STATE Solve for $\mathbf{G}_t$ in Eq.~(\ref{homo_obj}) given $\mathbf{B}_{t-1}$
            \STATE Compute $\mathbf{L}_t$ using $\mathbf{G}_t$
            \STATE $\mathbf{B}_t = (\mathbf{I} + \theta_2 \mathbf{L}_t )^{-1} (\mathbf{I} - \beta \mathbf{G}_t)\mathbf{A}$
            \STATE $\Delta = |h(\mathbf{G}_{t}, \mathbf{B}_{t}) - h(\mathbf{G}_{t - 1}, \mathbf{B}_{t - 1})|~~(\text{for}~~t>1)$
            \STATE $t = t+1$
            \ENDIF\label{euclidendwhile}
            \STATE \textbf{return:} $\mathbf{G} = \mathbf{G}_t, \mathbf{B} = \mathbf{B}_t$.
        \end{algorithmic}
    \end{algorithm}

    \section{Experiments on Synthetic Data}
    In this section, we evaluate the performance of the proposed learning frameworks on synthetic networks that follow three types of random graph models, i.e., the Erdős–Rényi (ER), the Watts-Strogatz (WS), and the Barabási-Albert (BA) models.
    In the ER graph, an edge is created with a probability of $p=0.2$ independently from all other possible edges. 
    In the WS graph, we set the average degree of the vertices to be $k=\lfloor \text{log}_2(N) \rfloor$, with a probability of $p=0.2$ for the random rewiring process.
    Finally, in the BA graph, we add $m=1$ new node at each time by connecting it to an existing node in the graph via preferential attachment. 
    All the graphs have $N=20$ vertices in our experiments.
    {Once the graphs are constructed, we compute $\beta>0$ such that the spectral radius, $\rho(\beta \mathbf{G})$, varies between 0 and 1 hence satisfying the assumption in Section~\ref{game}.}
    
    We adopt two different settings, one for generating the independent marginal benefits $\mathbf{b}$ and the other for the homophilous $\mathbf{b}$. In the independent case of Section~\ref{algo1}, for each game, we generate realizations by considering $\mathbf{b} \sim \mathcal{N}(\mathbf{0}, \mathbf{I})$. In the homophilous setting of Section~\ref{algo2}, {we generate realizations by considering $\mathbf{b} \sim \mathcal{N}(\mathbf{0}, \mathbf{L}^{\dagger})$, where $\mathbf{L}^{\dagger}$ is the Moore-Penrose pseudoinverse of the groundtruth graph Laplacian $\mathbf{L}$.
    In both cases we further add Gaussian noise $\boldsymbol{\epsilon} \sim \mathcal{N}(\mathbf{0}, \frac{1}{10}\mathbf{I})$ to the simulated marginal benefits.} 
    Now, given $\mathbf{b}$ and $\beta$, we 
    compute the players' Nash equilibrium action $\mathbf{a}$ according to Eq.~(\ref{equi}). 
    We consider $K=50$ games for each of which we generate the action $\mathbf{a}$.
    
    We apply Algorithm~\ref{alg:non-smooth} and Algorithm~\ref{alg:smooth} to the respective settings to infer graph structures and compare against the groundtruth ones in a scenario of binary classification, i.e., either there exists an edge between $i$ and $j$ (positive case), or not (negative case).
    {Since the ratio of positive cases is small for all the three types of graphs,} we use the area under the curve (AUC) for the evaluation of the learning performance. 
    We compare our algorithms with two baseline methods {for inferring graph structures given data observations:} the sample correlation and the regularized graphical Lasso in \citet{lake2010discovering}. 
    {In the former case we consider the correlations between each pair of variables as "edge weights" in a learned graph,
    while in the latter case a graph adjacency matrix is computed as in our algorithms.}
    
    {Notice that in the synthetic experiments, we focus on the case of strategic complements, i.e., $\beta>0$, to facilitate a fair comparison with the two baselines that only apply to this case. Our methods therefore also have the unique advantage of dealing with the case of strategic substitutes, i.e., $\beta<0$}.
    
    \begin{figure*}[t]
        \centering
        {\includegraphics[width=1.0\linewidth]{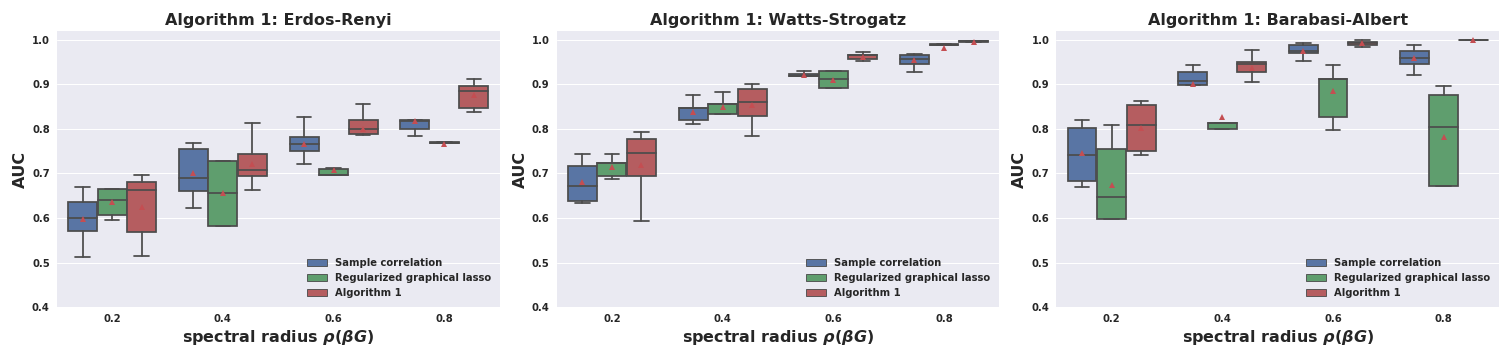}}\\
        {\includegraphics[width=1.0\linewidth]{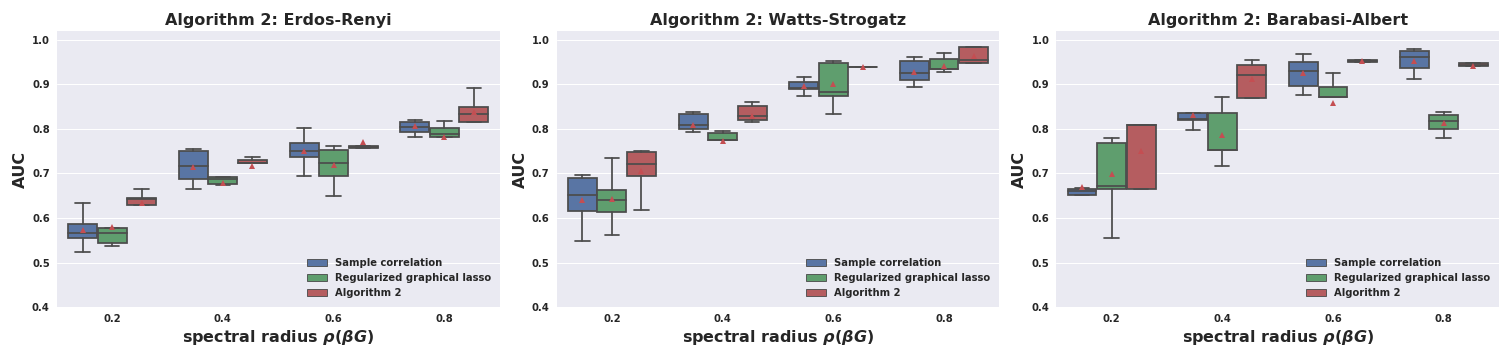}}
        \vspace{-0.4cm}
        \caption{Performance of the proposed algorithm and baselines in the setting of independent (top) and homophilous (bottom) marginal benefits. The red triangle, the middle line, lower and upper boundaries of the box (interquartile range or IQR) correspond to mean, median, and 25/75 percentile of the data, respectively. The lower and upper whiskers extend maximally 1.5 times of IQR from 25 percentile downwards and 75 percentile upwards, respectively.}
        \label{model_compare}
    \end{figure*}
    
    \subsection{Comparison of Learning Performance}
    The performance of the three methods in comparison is shown in Fig.~\ref{model_compare} (top) for the case of independent marginal benefits. For Algorithm~\ref{alg:non-smooth} and regularized graphical Lasso, we report the results using the parameter values that give the best average performance over 20 randomly generated graph instances\footnote{Analysis of robustness of performance against regularization parameters is presented in Supplementary Material.}.
    First, we see that the performance of all the three methods increases with the spectral radius $\rho(\beta \mathbf{G})$ for the majority of the cases. This pattern indicates that stronger strategic dependencies between actions of potential neighbors reveal more information about the existence of the corresponding links. Indeed, as $\rho(\beta \mathbf{G})$ increases, the action matrix $\mathbf{A}$ contains more information about the graph structure as explained in Section~\ref{sec:form1}.
    Second, the performance of the proposed Algorithm~\ref{alg:non-smooth} 
    generally outperforms the two baselines in terms of recovering the locations of the edges of the groundtruth.
    Notice that for regularized graphical Lasso, the performance drops with a larger value for $\rho(\beta \mathbf{G})$. One possible explanation is that, as $\rho(\beta \mathbf{G})$ becomes close to 1, 
    {the smallest eigenvalue of $\mathbf{I} - \beta \mathbf{G}$ approaches 0 resulting in a large ratio between the smallest and the largest eigenvalues of the empirical covariance of $\mathbf{a}$,}
    which may lead to inaccurate estimation of the precision matrix in the graphical Lasso.
    In comparison, our method does not seem to be affected by such phenomenon.
    Finally, the performance of all the methods for the WS and BA graphs is generally better than that of the ER graphs, possibly because there exists more structural information in the former models than the latter.

    
    
    The same results for the case of homophilous marginal benefits are shown in Fig.~\ref{model_compare} (bottom). We observe the same increase in performance as $\rho(\beta \mathbf{G})$ increases for all the three methods, as well as the drop in performance towards large $\rho(\beta \mathbf{G})$ for regularized graphical Lasso. The proposed Algorithm~\ref{alg:smooth} generally achieves superior performance in this scenario, which is expected due to the way the observations $\mathbf{A}$ are generated taking into account the regularization term in the objective in Eq.~(\ref{homo_obj}) that enforces homophily.

    \subsection{Learning Performance with Respect to Different Factors in Network Games}
    \label{sec:factor}
    {We now examine the performance of Algorithm~\ref{alg:smooth} with respect to a number of factors, including the number of games, the noise intensity, the structure of the groundtruth network, and the strength of the homophily effect in marginal benefits (in Supplementary Material).} {The same results for Algorithm~\ref{alg:non-smooth} are presented in Supplementary Material.}
    
    \textbf{Number of games.}
    We are first interested in understanding
    the influence of the number of games $K$ on the learning performance. 
    In the following and all subsequent analyses, we choose {$\rho(\beta \mathbf{G}) = 0.6$,} and fix the parameters 
    in Algorithm~\ref{alg:smooth} to be the ones that lead to the best learning performance. 
    In Fig.~\ref{factor-smooth2} (top), we vary the number of games and evaluate its effect on the performance. 
    {We see that in general, the performance of the algorithm increases, as more observations become available. The benefit is least obvious for the ER graph, 
    suggesting that adding more observations does not help as much in improving the performance in this case when the edges in the graph appear more randomly.}
    
        \begin{figure*}[t]
            \centering
            \includegraphics[width=1.0\linewidth]{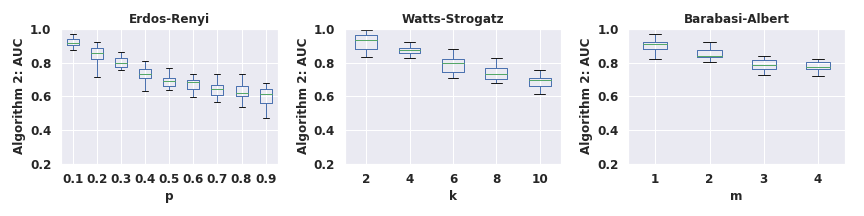}
            \vspace{-0.6cm}
            \caption{Performance of Algorithm~\ref{alg:smooth} versus 
            structural properties of the network.}
            \label{factor-smooth}
        \end{figure*}
        
    \textbf{Noise intensity {in the marginal benefits.}}
    We now analyze the robustness of the result against noise intensity {in the marginal benefits.} 
    {With more noise in the marginal benefits, the observed actions $\mathbf{A}$ becomes noisier as well, hence possibly affecting the learning performance.}
    As shown in Fig.~\ref{factor-smooth2} (bottom),
    the learning performance generally decays as the intensity of noise increases, which is expected.
    The performance of the model is relatively stable until the standard deviation of the noise becomes larger than 1.

        
    \textbf{Network structure.}
    The random graphs used in our experiments have parameters that may affect the performance of the proposed algorithms. We, therefore, analyze the effect of $p$ in the ER, $k$ in the WS, and $m$ in the BA graphs on the learning performance of the proposed algorithm.
    {The larger these parameters, the higher the edge density in these random graph models.}
    As shown in Fig.~\ref{factor-smooth}, 
    the density of edges has a substantial effect on the learning performance for all the networks, i.e., the denser the edges, the worse the performance. One possible explanation is that, in a sparse network 
    the correlations between individuals' actions might contain more accurate information about the existence of dependencies hence edges between them, while in a dense network the influence from one neighbor is often mingled with that from another, which makes it more challenging to uncover pairwise dependencies.
    
    

    \subsection{Learning the Marginal Benefits}
    {In our framework, we jointly infer the graph structure and the marginal benefits of the players.
    This is one of the main advantages of our algorithms,} since the inference of marginal benefits can be critical for targeting strategies and interventions \cite{galeotti2017targeting}. 
    To test the performance of learning marginal benefits, for each random graph model, we generate a network with 20 nodes and simulate 50 games with $\rho(\beta \mathbf{G}) = 0.6$, for both independent and homophilous marginal benefits. We repeat this process for 30 times, 
    {and report the average performance of learning the marginal benefits} in Table \ref{tab:learned_B}. The performance is measured in terms of the coefficients of determination ($R^2$), by treating the groundtruth and learned marginal benefits (both in vectorized form) as dependent and independent variables, respectively.
{As we can see, in both cases the $R^2$ values are above 0.9, which indicates that the learned marginal benefits are very similar to the groundtrith ones.} 

\begin{table}[h]
\centering
\caption{Performance ($R^2$) of learning marginal benefits.}
\scalebox{0.8}{
\begin{tabular}{|l|c|c|c|c|}
\hline
\multirow{2}{*}{} & \multicolumn{2}{c|}{Algorithm~\ref{alg:non-smooth}}    & \multicolumn{2}{c|}{Algorithm~\ref{alg:smooth}} \\ \cline{2-5} 
                  & \multicolumn{1}{c|}{mean} & \multicolumn{1}{c|}{std} & mean                  & std                  \\ \hline
ER graph       & 0.959                    & 0.005                    & 0.982                & 0.002                \\ \hline
WS graph    & 0.955                    & 0.007                    & 0.921                & 0.010                \\ \hline
BA graph   & 0.937                    & 0.008                    & 0.909                & 0.010                \\ \hline
\end{tabular}}
\vspace{-0.2cm}
\label{tab:learned_B}
\end{table}

    \section{Experiments on Real-World Data}
    The strategic interactions between players in real-world situations may follow the formulation of network games. Given this broad assumption,
    we present three examples of inferring the network structure in quadratic games in practical scenarios, e.g., the inference of social, trade, and political networks.
    
    \subsection{{Social Network}}
    {We first consider inferring a social network between households in a village in rural India} \cite{banerjee2013diffusion}. In particular, {following the setting in \citet{banerjee2013diffusion}, we consider the actions of each household as choosing the number of rooms, beds, and other facilities in their houses.
    The assumption is that there may exist strategic interactions between these households regarding constructing such facilities. 
    In particular, when deciding to adopt new technologies or innovations, people have an incentive to conform to the social norms they perceive \cite{young2009innovation, montanari2010spread}, which are formed by the decisions made by their neighbors. For example, if neighbors adopt a specific facility, villagers tend to gain higher payoff after adopting the same facility by complying with social norms (i.e., strategic complements).
    
    We consider each action as a strategy in a quadratic game, and we have 31 games with discrete actions made by 182 households.}
    We then apply the proposed algorithms to infer the relationships 
    {between these households, and compare against a groundtruth network of self-reported friendship.}
    {Since we do not observe $\beta$, we treat it as a hyperparameter,
    and tune it within the range of $\beta \in [-3, 3]$. }
    It can be seen from Table~\ref{tab:exp12} that {both of the proposed methods} outperform regularized graphical Lasso by about 2.5\% and sample correlation by about 10.7\%\footnote{The improvement is calculated by the absolute improvement in AUC normalized by the room for improvement. 
    The best performance of Algorithm~\ref{alg:non-smooth} is obtained with $\beta = 0.1$, $\theta_1 = 2^{-8.5}$, and $\theta_2 = 2^{1}$, while that of Algorithm~\ref{alg:smooth} is obtained with $\beta = 2.6$, $\theta_1 = 2^{7}$, and $\theta_2 = 2^{-5.5}$. The positive sign of $\beta$ in both cases indicates a strategic complement relationship between the households, which is consistent with our hypothesis.}, indicating that they can recover a social network structure closer to the groundtruth.
    

    \subsection{{Trade Network}}
    We next consider inferring a global trade network. Specifically, we consider the overall trading activities of 235 countries on 96 export products and 96 import products in the year 2008 as our observed actions\footnote{Data can be accessed via \url{https://atlas.media.mit.edu/en/resources/data/}. The trading activities are classified by the 2002 edition of the HS (Harmonized System).}. This leads to 192 games {(for both import and export actions)} played by 235 agents {(countries)}. By applying the proposed algorithms, we infer the relationships among nations regarding their strategic trading decisions and compare against a groundtruth, which is the trading network in year 2002\footnote{The trading network from previous years provides a foundation for nations to make decisions and thus can be thought of as a groundtruth. The year 2002 is the latest year before 2008 for which trading data are available.}. 
    In constructing the groundtruth, we consider the edge weight between each pair of nations as the logarithmic of the total amount of trades (import plus export) between the two nations.
    
    
    In the groundtruth trade network, each nation is connected with the ones with which it traded in 2002. 
    This implies that the nation has different demand and supply compared to its neighbors, and their import and export actions tend to be different in the near future. 
    Therefore, we expect a strategic substitute relationship between the nations when looking at their import and export activities in 2008.
    
    We tune $\beta$ within the range of $\beta \in [-1, 1]$. Table~\ref{tab:exp12} shows that Algorithm~\ref{alg:non-smooth} and Algorithms 2 outperform regularized graphical Lasso by 12.09\% and 24.85\%, respectively\footnote{The best performance of Algorithm~\ref{alg:non-smooth} is obtained with $\beta = -0.6$, $\theta_1 = 2^{1}$, and $\theta_2 = 2^{-10}$, and that of Algorithm~\ref{alg:smooth} is obtained with $\beta = -0.7$, $\theta_1 = 2^{11.5}$, and $\theta_2 = 2^{-15.5}$. The negative sign of $\beta$ in both cases indicates a strategic substitute relationship between the nations, which is consistent with our hypothesis.}. 
    The larger performance gain in this case is due to the fact that both sample correlation and regularized graphical Lasso are suitable only for strategic complement and not strategic substitute relationships.
    Furthermore, Algorithm~\ref{alg:smooth} performs better than Algorithm~\ref{alg:non-smooth} in this example, which implies a homophilous distribution of marginal benefits across neighboring nations.


    \begin{table}[h]
    \caption{Performance (AUC) of learning {the structure of} the social network and the trade network.}
     \centering
     \scalebox{0.8}{
        \begin{tabular}{|l|c|c|}
        \hline
                                    & Social network & Trade network\\ \hline
        Sample correlation          & 0.525 & 0.523\\ \hline
        Regularized graphical Lasso & 0.564 & 0.570\\ \hline
        Algorithm~\ref{alg:non-smooth}                 & 0.575 & 0.622\\ \hline 
        Algorithm~\ref{alg:smooth}                 & 0.576 & 0.677\\ \hline 
      \end{tabular}}
      \label{tab:exp12}
    \end{table}


    
    


    \begin{figure*}[t]
    \centering
    \includegraphics[width=0.4\linewidth]{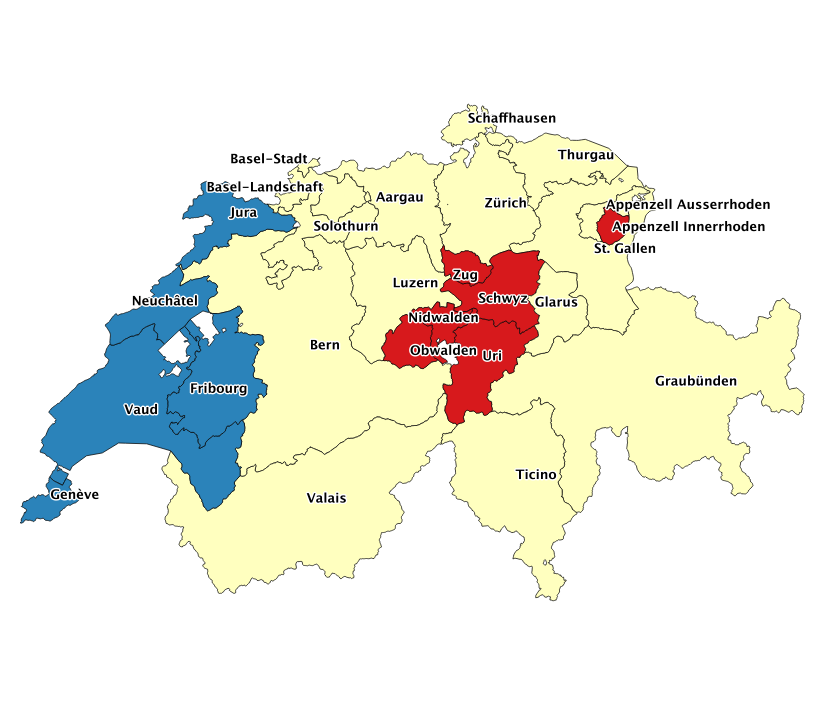}
    \includegraphics[width=0.4\linewidth]{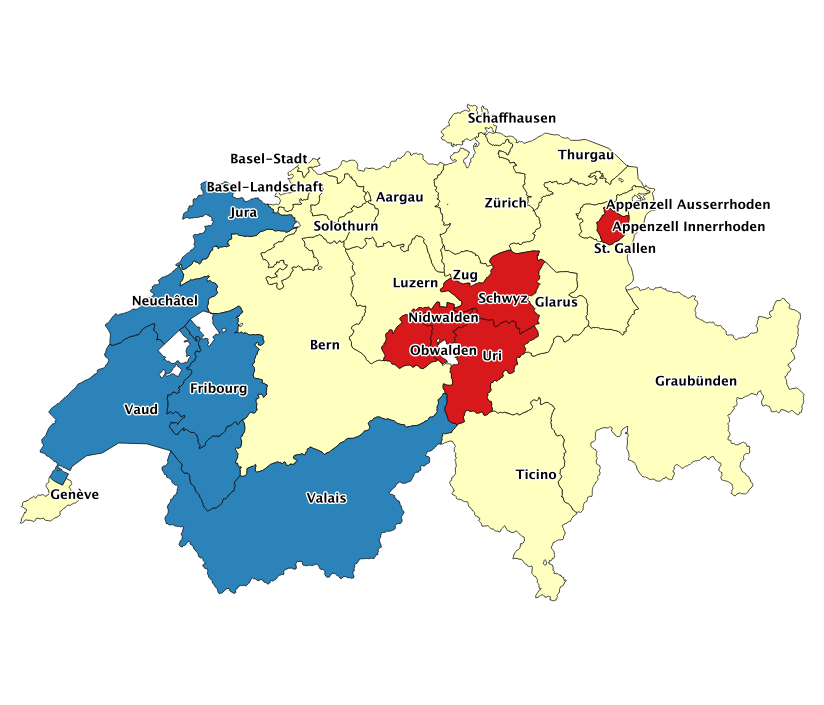}
    \caption{Clustering of Swiss cantons based on the political network learned by Algorithm~\ref{alg:non-smooth} (left) and Algorithm~\ref{alg:smooth} (right).}
    \label{votes}
    \end{figure*}

\subsection{{Political Network}}
    The third real-world example we considered is {the inference of the relationship between the cantons in Switzerland in terms of their political preference.}
    To this end, we consider voting statistics from the national referendums for 37 federal initiatives in Switzerland between 2008 and 2012\footnote{The voting statistics were obtained via \url{http://www.swissvotes.ch.}}. Specifically, we consider the percentage of voters supporting each initiative in the 26 Swiss cantons as the observed actions. This leads to 37 games {(initiatives)} played by 26 agents {(cantons).} 
    By applying the proposed algorithms, we infer a network that captures the {strategic} political relationship between these cantons reflected by their votes in the national referendums\footnote{We tune $\beta$ within the range of $[-1, 1]$. For Algorithm \ref{alg:non-smooth} we report results with $\beta = 0.6$, $\theta_1 = 2^{-6.2}$, and $\theta_3 = 2^{-1.65}$. For Algorithm \ref{alg:smooth} we report results with $\beta = 0.67$, $\theta_1 = 2^2$, and $\theta_2= 2^3$. The positive sign of $\beta$ in both cases indicates a strategic complement relationship between the cantons.}.
  
    {Unlike the previous examples, it is more difficult to define a groundtruth network in this case. Instead, we apply spectral clustering \cite{Luxburg07} to the learned network and interpret the obtained clusters of cantons.}
    The three-cluster partition of the networks learned by Algorithm~\ref{alg:non-smooth} and Algorithm~\ref{alg:smooth} are presented in Fig.~\ref{votes}. As we can see, the clusters obtained in the two cases are largely consistent, with the blue and yellow clusters generally corresponding to the French-speaking and German-speaking cantons, respectively. The red cluster, in both cases, contains the five cantons of Uri, Schwyz, Nidwalden, Obwalden and Appenzell Innerrhoden, which are all considered among the most conservative ones in Switzerland.
    This demonstrates that the learned networks are able to capture the strategic dependence between cantons within the same cluster, which tend to vote similarly in national referendums.
    

    \section{Discussion}
    In this paper, we have proposed two novel learning frameworks for joint inference of graph structure and individual marginal benefits for a broad class of network games, i.e., games with linear-quadratic payoffs.
    We believe that the present paper may shed light on the understanding of network games (in particular those with linear-quadratic payoffs), and contribute to the vibrant literature of learning hidden relationships from data observations.

    The proposed approaches can benefit a wide range of practical scenarios. For instance, the learned graph, which captures the strategic interactions between the players, may be used for detecting communities formed by the players \cite{fortunato2010community}. This can, in turn, be used for purposes such as stratification. Another use case is to compute centrality measures of the nodes in the network, which may help in designing efficient targeting strategies in marketing scenarios \cite{leng2018contextual}. Finally, the joint inference of the graph and the marginal benefits can help a central planner who wishes to design intervention mechanisms achieve specific planning objectives. One such objective could be the maximization of the total payoffs of all players, which can be done by adjusting, according to the network topology, the marginal benefits via incentivization \cite{galeotti2017targeting}. Another objective could be the reduction of inequality between the players in terms of their payoffs, which can be done by adjusting network topology via encouraging the formation of certain new relationships.

    There remain many interesting directions to explore. For example, building upon the promising empirical results presented in this paper, it would be important to study the theoretical guarantees of the proposed algorithms in recovering the graph structure.
    It would also be interesting to consider graph inference given partial or incomplete observations of the actions, especially in the case where it is costly to observe the actions of all the network players, or consider a setting where the underlying relationships between the players may evolve over time, which can be modeled by dynamic graph topologies. Finally, the inference framework might need to be adapted accordingly for network games of different payoff functions. In this sense, it would be very interesting to investigate the possibility of inferring the graph structure in a purely date-driven fashion via a neural network, without the explicit knowledge of the payoff function. We leave these studies as future work.

\section*{Acknowledgement}
The authors would like to thank Georgios Stathopoulos, Dorina Thanou, and Ye Pu for helpful discussions about the optimization problems in the paper.

\bibliographystyle{icml2020}
\bibliography{merged_file}

\clearpage
\beginsupplement
\section*{Supplementary Material}
\subsection*{Robustness against Regularization Parameters}

\begin{figure*}[t]
        \centering
            \includegraphics[width=0.9\linewidth]{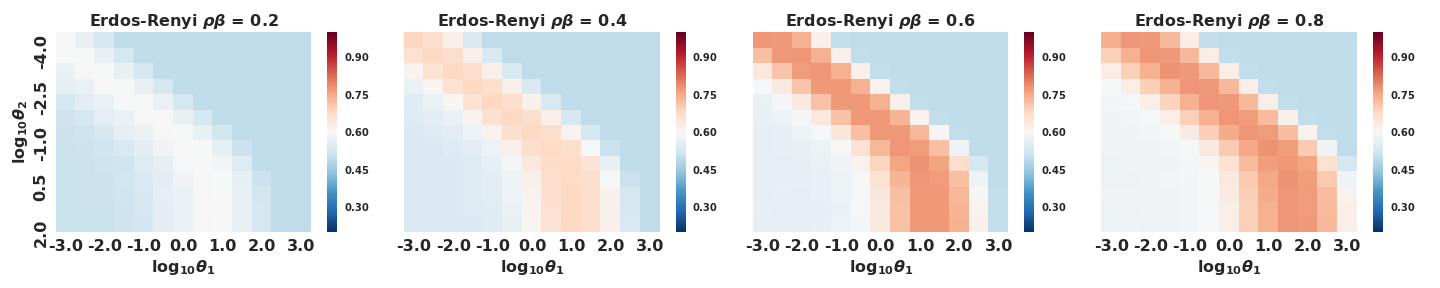}
            \includegraphics[width=0.9\linewidth]{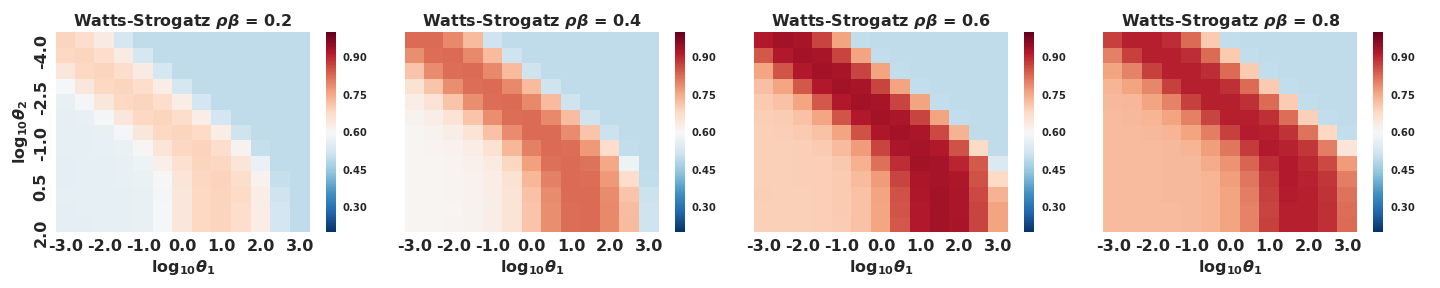}
            \includegraphics[width=0.9\linewidth]{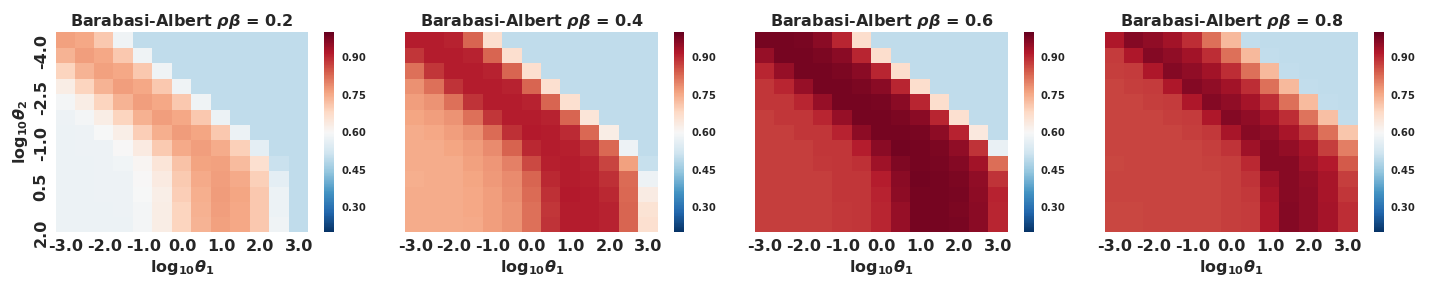}
        \caption{Performance (AUC) of Algorithm~\ref{alg:non-smooth} {with respect to} $\rho(\beta \mathbf{G})$, $\theta_1$, and $\theta_2$.}
        \label{non_smoothing}
    \end{figure*}

    We analyze the robustness of the performance of Algorithm~\ref{alg:non-smooth} against the regularization parameter $\theta_1$ in Eq.~(\ref{obj_min1}),
    and the results averaged over 20 random graph instances are presented in Fig.~\ref{non_smoothing}.
    In general, in addition to the effect of $\rho(\beta \mathbf{G})$ discussed in the main text,
    {we see a consistent pattern across the three graph models that link the values of $\theta_1$ and  $\theta_2$ to the learning performance. Specifically, when $\theta_1$ is smaller than around $10^{2}$, there is a region where a certain ratio of $\theta_1$ to $\theta_2$ leads to optimal performance, suggesting that in this case, the second and third terms are the dominating factors in the optimization of Eq.~(\ref{obj_min1}). A phase transition takes place when $\theta_1$ is larger than $10^{2}$, where the performance becomes largely constant. The reason behind this behavior is as follows. When $\theta_1$ increases, the Frobenius norm of $\mathbf{G}$ in the objective function of Eq.~(\ref{obj_min1}) tends to be small. Given a fixed element-wise $L^1$-norm of $\mathbf{G}$, this leads to a more uniform distribution of the off-diagonal entries. When $\theta_1$ is large enough, the edge weights become almost the same, leading to a constant AUC measure.}
    
\begin{figure*}
        \centering    
            \includegraphics[width=0.9\linewidth]{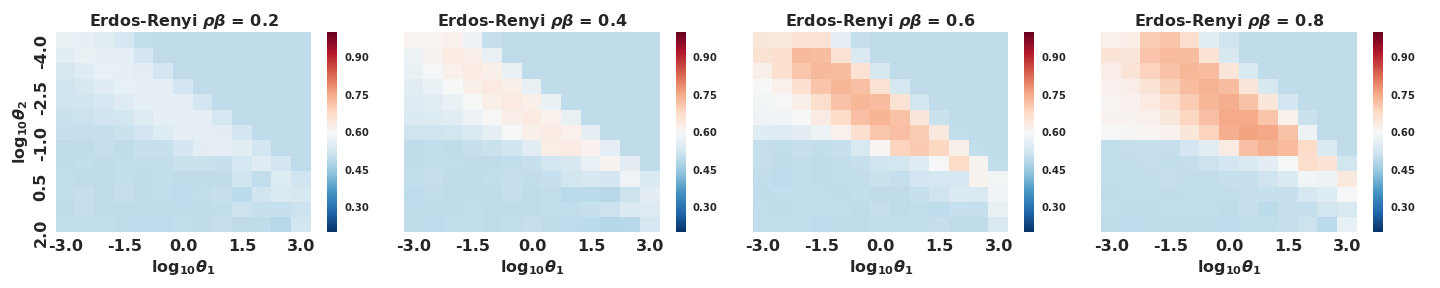}
            \includegraphics[width=0.9\linewidth]{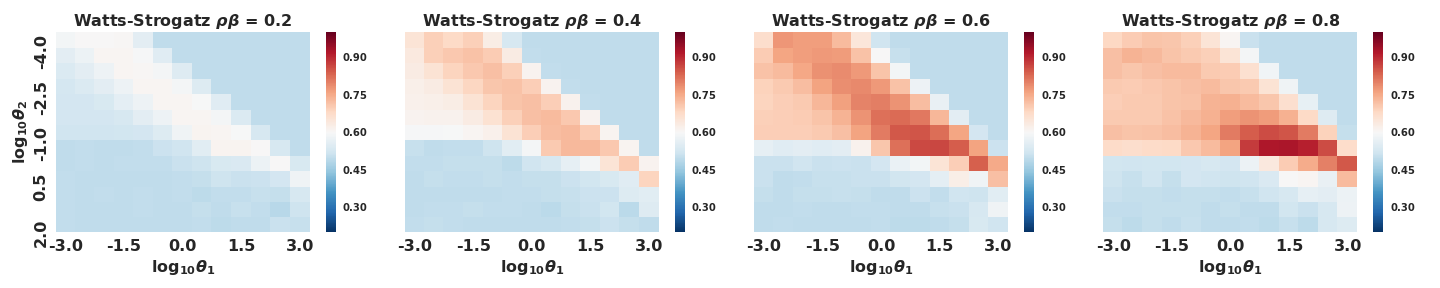}
            \includegraphics[width=0.9\linewidth]{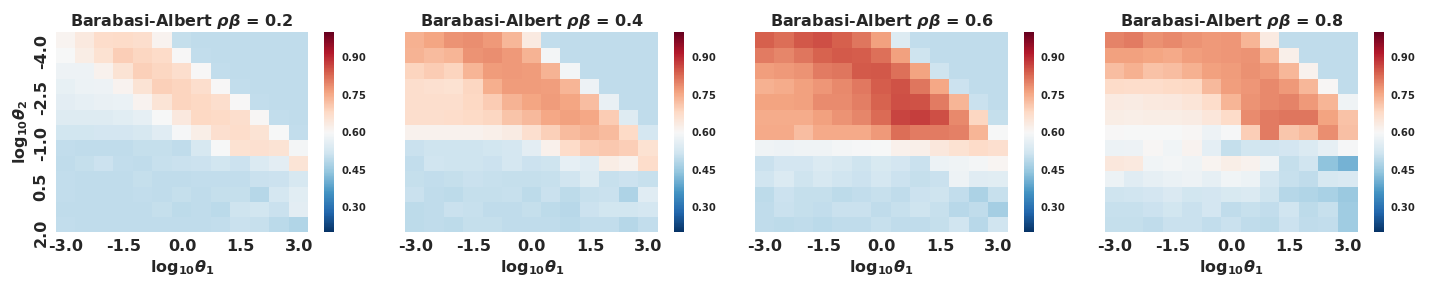}
            \caption{Performance (AUC) of Algorithm~\ref{alg:smooth}  {with respect to} $\rho(\beta \mathbf{G})$, $\theta_1$, and $\theta_2$.}
        \label{smoothing}
    \end{figure*}

    Similarly, we present in Fig.~\ref{smoothing} the performance of Algorithm~\ref{alg:smooth} with respect to different values of $\theta_1$ and $\theta_2$ in Eq.~(\ref{homo_obj}). 
    {We see that the patterns are generally consistent with that in Fig.~\ref{non_smoothing}, with one noticeable difference being that there also seems to be a phase transition taking place around the value of $10^{-1}$ for $\theta_2$. One possible explanation for this behavior is that, when $\theta_2$ is large enough, the trace term in the objective function of Eq.~(\ref{homo_obj}) tends to be small, making the resulting graph with fewer edges but with larger weights. This contributes to an AUC score that is mostly constant.}
    
\subsection*{Performance of Algorithm~\ref{alg:smooth} with respect to Number of Games and Noise Intensity in Marginal Benefits}

\begin{figure*}[t]
            \centering
            {\includegraphics[width=0.8\linewidth]{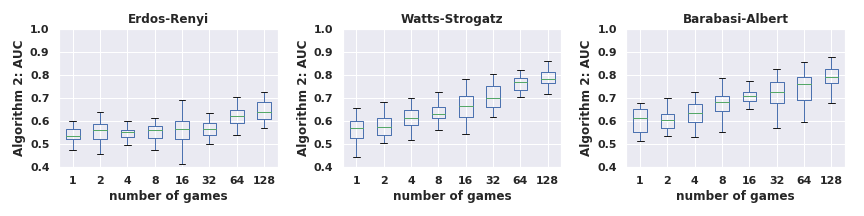}} \\
            {\includegraphics[width=0.8\linewidth]{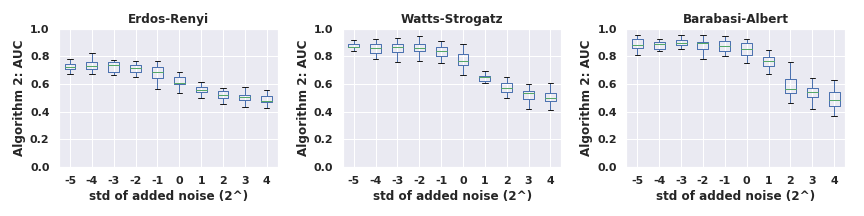}} 
            \caption{Performance of Algorithm~\ref{alg:smooth} versus number of games (top) and noise intensity in marginal benefits (bottom).}
            \label{factor-smooth2}
        \end{figure*}

The performance of Algorithm~\ref{alg:smooth} with respect to the number of games and noise intensity in marginal benefits analysed in Section~\ref{sec:factor} is presented in Fig.~\ref{factor-smooth2}.

\subsection*{Performance of Algorithm~\ref{alg:smooth} with respect to Strength of Homophily Effect}

\begin{figure*}[t]
        \centering
        \includegraphics[width=0.9\linewidth]{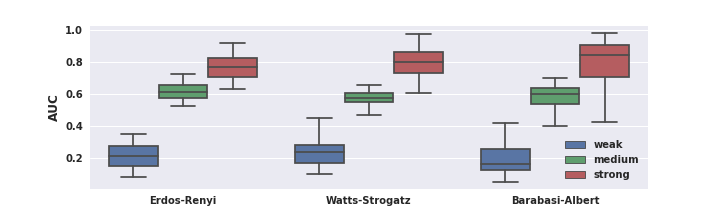}
        \caption{Performance of Algorithm~\ref{alg:smooth} versus strength of homophily in the marginal benefits. 
        }
        \label{homophily_plot}
    \end{figure*}

    We analyze the influence of the strength of homophily on the learning performance of Algorithm~\ref{alg:smooth}. We consider three scenarios, i.e., weak, medium and strong homophily effect. To this end, we generate the marginal benefits $\mathbf{b}$ as linear combinations of the eigenvectors corresponding to the $1^\text{st}$-$5^\text{th}$, $6^\text{th}$-$10^\text{th}$, and $11^\text{th}$-$15^{\text{th}}$ smallest eigenvalues of the graph Laplacian.
    {Due to the properties of the eigenvectors, these three sets lead to different quantities for the Laplacian quadratic form, hence corresponding to weak, medium and strong homophily effect, respectively.}
    {Notice that the presence of the homophily effect in $\mathbf{B}$ tends to imply homophily in $\mathbf{A}$ for the following reason. Regardless of the characteristics of the game, a higher marginal benefit $\mathbf{b}$ is more likely to incentivize higher activity level $\mathbf{a}$ due to the first term of the payoff function in Eq.~(\ref{utility}). Therefore, homophily in $\mathbf{B}$ tends to lead to homophily in $\mathbf{A}$, hence revealing more information about the graph structure.}
    As shown in Fig.~\ref{homophily_plot}, for all the three types of networks, the stronger the homophily in the marginal benefits, the better the learning performance. 
    
    
\subsection*{Results in Section~\ref{sec:factor} for Algorithm~\ref{alg:non-smooth}}

\begin{figure*}[t]
            \centering
            {\includegraphics[width=0.9\linewidth]{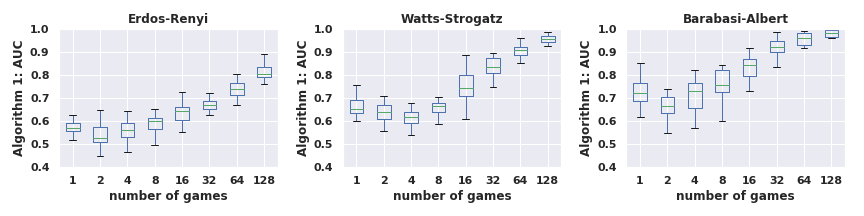}} \\
            {\includegraphics[width=0.9\linewidth]{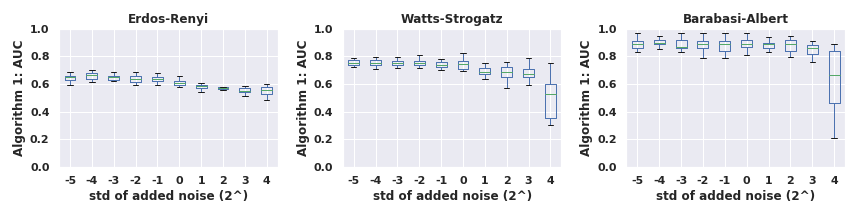}} \\
            \includegraphics[width=0.9\linewidth]{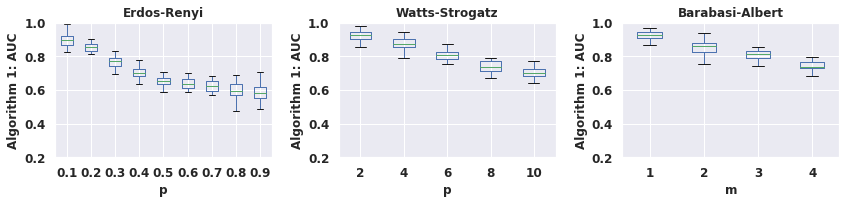}
            \caption{Performance of Algorithm~\ref{alg:non-smooth} versus number of games (top), noise intensity in marginal benefits (middle), and structural properties of the network (bottom).}
            \label{factor-nonsmooth}
        \end{figure*}

The performance of Algorithm~\ref{alg:non-smooth} with respect to the factors analysed in Section~\ref{sec:factor} is presented in Fig.~\ref{factor-nonsmooth}.

\end{document}